\documentclass[twocolumn,twoside]{revtex4}
\usepackage{graphicx}
\usepackage{fancyhdr}
\begin{document}

\title{Muon Campus at Fermilab}

%

\author{S.~Ganguly\\
}
\affiliation{Fermi National Accelerator Laboratory, Batavia, IL, USA, 60510}
\begin{abstract}
The Muon Campus at Fermilab provides world class accelerator infrastructure supporting the next generation intensity frontier experiments. The anti-proton source from the Tevatron era was converted to the present day Muon Campus at the end of the collider program in 2011. Currently, the Muon Campus delivers highly polarized muon beams to the Muon g-2 experiment and in a few years, it will provide muon beams to the Mu2e experiment. The Muon g-2 experiment is a high precision test of the standard model; it is precise enough to measure QED, Weak, and QCD Standard Model contributions.The Muon g-2 experiment looks for a significant deviation from the Standard Model whereas the Mu2e experiment looks for rare processes using high intensity muon beams. 
The Muon g-2 experiment has measured the anomalous magnetic moment of muons to an unprecedented 460 ppb precision with Run 1 data. The experiment has completed 5 runs, accumulating 19$\times$ Brookhaven E821 data set. 
The Mu2e experiment has completed the construction phase and first beam to the Diagnostic Absorber was run on April 14, 2022. 
Installation and commissioning of the Mu2e experiment will begin in 2023 and beam commissioning and physics run is currently planned for 2026. 
\end{abstract}
\maketitle
\thispagestyle{fancy}

\section{Introduction}
The Muon Campus at Fermilab \cite{muondept} was built after the collider program ended in 2011. The former anti-proton source beamlines and rings were repurposed, redesigned, rebuilt and upgraded to support the Muon g-2 and the Mu2e experiments.
In transforming the former antiproton source to the present day Muon Campus, several changes were made. The 120 GeV extraction from Main Injector for antiproton stacking was replaced by the 8 GeV extraction from the Recycler for the Muon Campus. The 120 GeV protons on target for the anti-proton production was replaced by the 8 GeV protons on target for muon production for the Muon g-2 experiment and the 8 GeV transport for the Mu2e experiment bypassing the Muon g-2 target. The antiproton accumulator ring was removed, and the Debuncher was reused as the Delivery Ring (DR) \cite{DR}. Two new beamlines were built to transport beam to the Muon g-2, and Mu2e experiments. The Muon Campus facility has been commissioned in 2017 and is now in operation phase for the Muon g-2 Experiment since 2018. A polarized beam of positive muons is injected into the experiment's storage ring \cite{storagering} with a vertical uniform magnetic field. The Muon Campus currently delivers $\sim 1\times$ Brookhaven E821 \cite{BNL} data set statistics per month to the experiment.
The Muon g-2 experiment at Fermilab has recently measured the muon anomalous magnetic moment to an unprecedented precision of 460 parts per billion \cite{gm2result} with the Run 1 dataset. 
The Mu2e experiment will improve the sensitivity on the search for the as-yet unobserved charged lepton flavor violation process of a neutrino-less conversion of a muon to an electron by four orders of magnitude. The first beam to the Mu2e Diagnostic Absorber was sent on April 14, 2022. The experiment will use an intense pulsed negative muon beam that will be crucial in background reduction. Installation and commissioning of the the Mu2e experiment will begin in 2023 after the Muon g-2 experiment ends and the experiment will need to collect about 3 years of physics data to reach a SES (single event sensitivity) of $3 \times 10^{17}$ \cite{mu2estatus} on the $\mu \to e$ conversion rate.
\section{MUON CAMPUS COMMISIONING}
\begin{figure}[h]
\centering
\includegraphics[width=76mm]{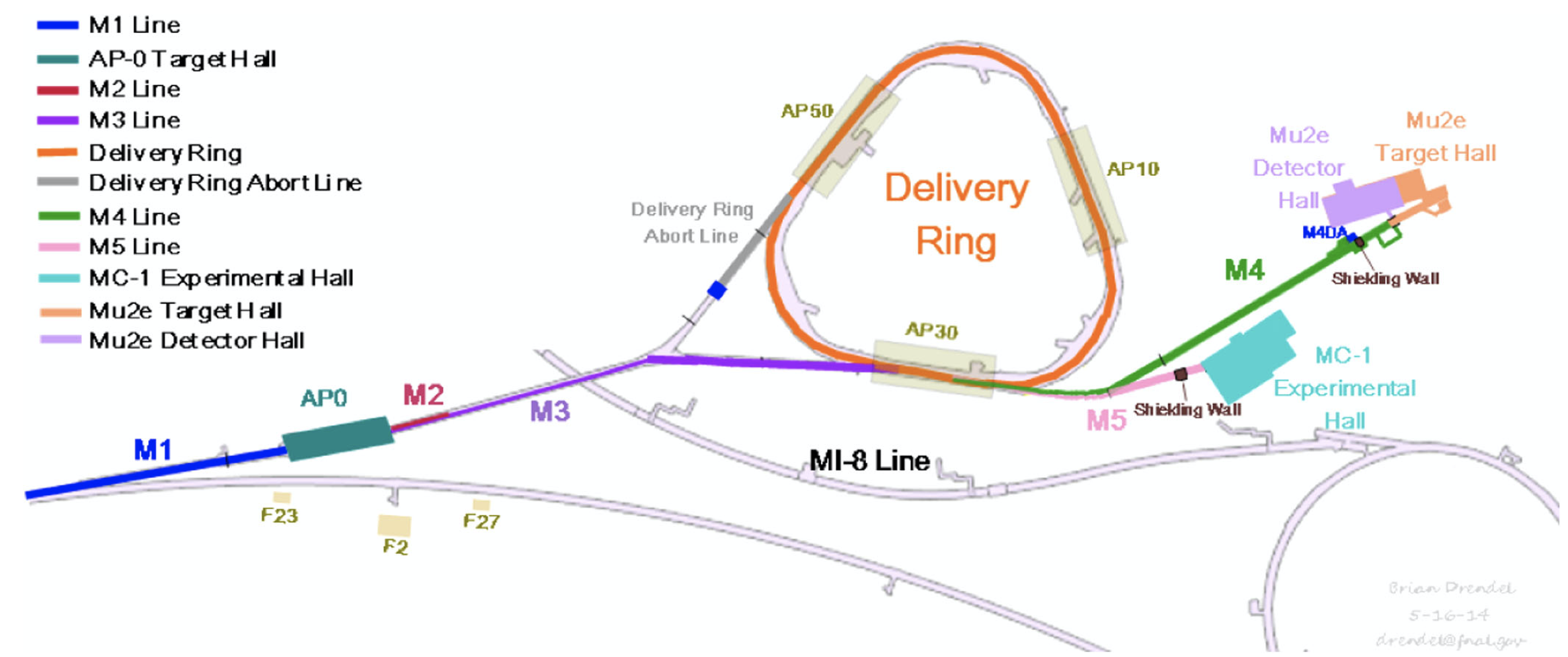}
\caption{A schematic representation of the Muon Campus at Fermilab.} \label{fig1}
\end{figure}

Figure \ref{fig1} shows an illustration of the Muon Campus at Fermilab that is currently being used by the Muon g-2 experiment \cite{diktys}. 
The first phase of the commissioning began in April, 2017 when a 8 GeV proton beam from the Recycler was sent to the Delivery Ring by bypassing the production target. Then the beam was extracted into the M4 line. The aim of the first phase of the commissioning was to test the beamline optics and the instrumentations with the primary proton beam. During the second phase of the commissioning, a secondary beam of 3.1 GeV was created by impinging primary proton beam on the target at AP0 target station. The secondary beam was sent straight through the DR and passed on to the M4/M5 beamlines and into the storage ring in the Mupon g-2 experimental hall. During the last phase, the secondary 3.1 GeV beam was created and was circulated around the DR to kick protons out after the fourth turn and the muon beam was sent to the storage ring via the M4/M5 lines. 
\section{Differences between The Muon g-2 experiment and the Mu2e experiment Beam Delivery}
\begin{figure}[h]
\centering
\includegraphics[width=76mm]{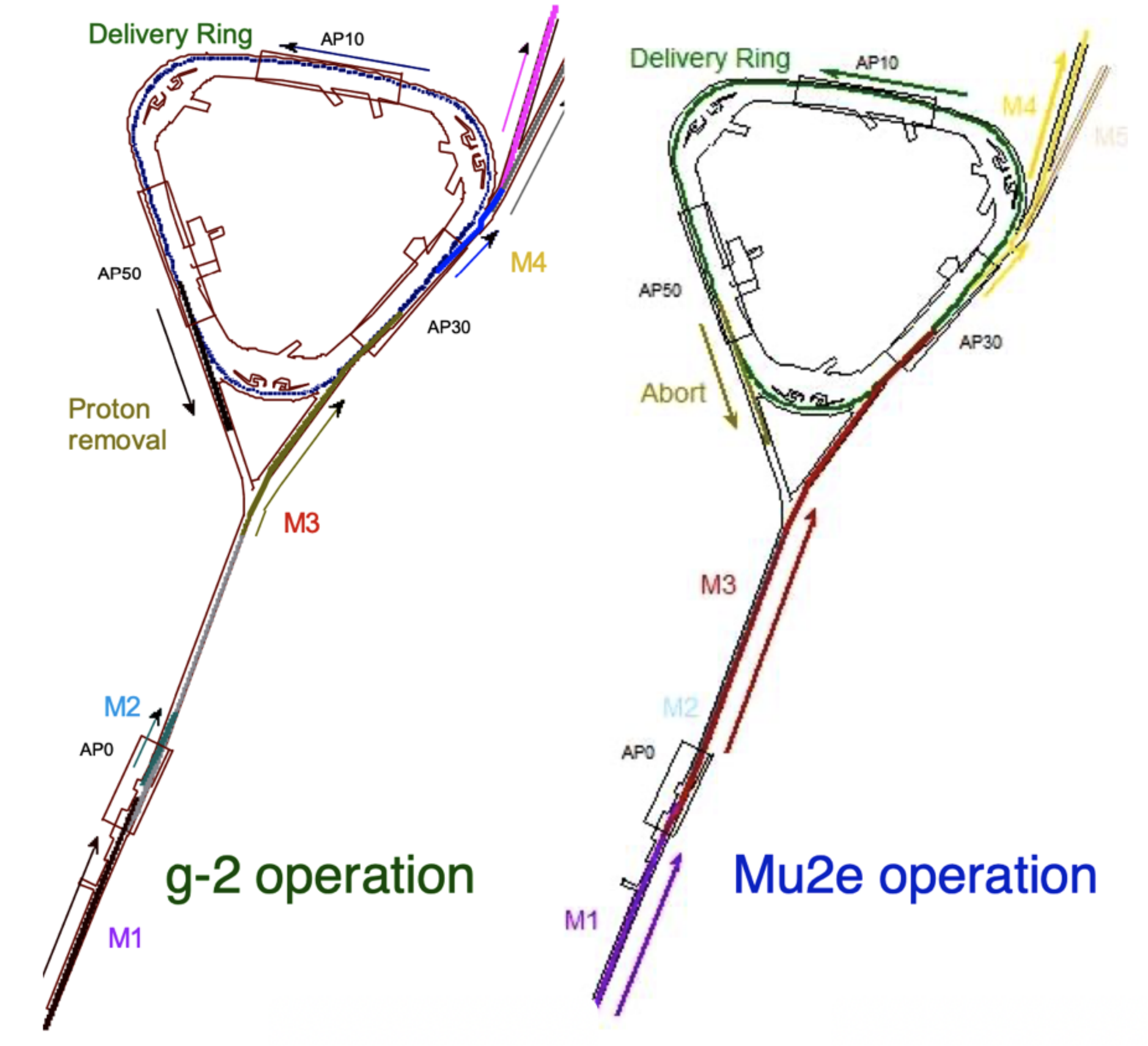}
\caption{Differences between Muon g-2 and Mu2e beam delivery.} \label{fig2}
\end{figure}

The 8.89 GeV/c proton beam from the Recycler ring is transported to the AP0 target station for both the Muon g-2 and the Mu2e experiments. A 3.1 GeV secondary beam is then transported through M2 and M3 beamlines to the Delivery Ring for the Muon g-2 experiment whereas for the Mu2e experiment, the AP0 target is bypassed and the proton beam is resonantly extracted from the DR over 43 ms into the M4 beamline and transported to the Mu2e production target. For the Muon g-2 experiment, beam takes 4 turns around the DR and then it is extracted into the M4 and M5 lines to the Muon g-2 storage ring over a single turn. The differences in the beam delivery for the two experiments are highlighted in Figure \ref{fig2}.
\section{Muon g-2 Beam Delivery Details}
Protons of 8 GeV kinetic energy from the Booster \cite{booster} will be manipulated to form four bunches with reduced intensity in each bunch in the Recycler. Each bunch is transported one at a time through the P1, P2, and M1 beamlines to a Inconel target \cite{jmorgan} at AP0. The protons impinging on the target produces a secondary beam composed of protons, pions, positrons, and deuterons. 
The secondaries are focused by a lithium lens \& momentum-selected by pulsed dipole magnet with a momentum of $3.11 GeV/c(\pm \sim 10\%)$. The secondaries travel through the M2 and M3 lines which have high magnet density with large aperture quadrupoles. These beamlines capture as many muons from forward pion decays as possible, with a momentum of 3.094 GeV/c. $95\%$ polarization of the muon is achieved along these beamlines. $70\%$ of the pions are expected to decay along M2-M3 lines before reaching the Delivery Ring. The secondary beam is then transferred to the DR. During injection, at the DR the beam composition is 89$\%$ protons, $8\%$ pions, $2\%$ muons. After 4 turns around the DR, only protons and muons remain. A kicker is used to remove the protons which are then sent to the beam dump. And the muon beam is then extracted to the experiment storage ring via the M5 beamline. 

\subsection{Recycler Ring 2.5 MHz Beam Rebunching}
The full intensity of a single Booster batch needs to be readjusted into four bunches, each bunch containing reduced intensity, for the Muon g-2 detectors to be able to accept a low number of muons in a given segment. 
\begin{figure}[h]
\centering
\includegraphics[width=76mm]{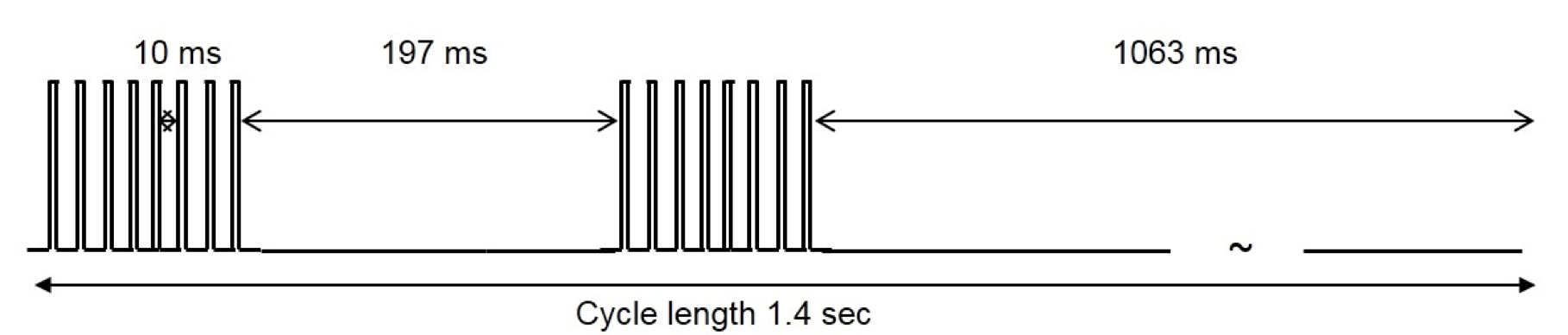}
\caption{Time structure of the beam pulses to the Muon g-2 experiment.} \label{fig3}
\end{figure}
The RF manipulations known as "Rebunching" \cite{rebunch} of a Booster batch containing $4 \times 10^{12}$ protons in the Recycler divides a batch into four bunches of $1 × 10^{12}$ protons each. Figure \ref{fig3} shows that the RF manipulation of the proton beam in the Recycler, allows the experiment to take four Booster batches or sixteen bunches in a 1.4 s cycle, where each bunch within a batch is separated by 10 ms, so that the muons have enough time to decay away with time-dilated muon lifetime in the lab frame being $\sim 64.4\mu$s. This also makes sure that there is enough time for data acquisition in the detector. The longitudinal width of each of the bunches is at most 120 ns as the muons take 149 ns to travel around the storage ring. 
The Rebunching process as shown in Figure \ref{fig4} begins when the first batch from Booster is injected into the Recycler at $t=0$. Each Booster batch contains eighty four 53 MHz bunches. The second batch is injected at $t = 200$. Rebunching ramp plays immediately after the second bunch is injected. After rebunching, eight 2.5 MHz bunches each containing $10^{12}$ protons are created that circulate in the Recycler ring. Initially, the Booster bunches are injected into matched buckets of 80 kV of 53 MHz RF. Then the 53 MHz voltage is switched off and the 2.5 MHz RF is turned on at 3 kV and then it is slowly ramped to 80 kV over 90 ms. The advantage of applying the adiabatic ramping is that it minimizes the possibility of bunch rotations in mismatched 53 MHz and 2.5 MHz RF buckets. The first 2.5 MHz bunch of protons are extracted to the Muon g-2 target at $t = 460$. The 2.5 MHz bunches are extracted successively every 10 ms. The rebunching of the Booster batches in the Recycler to create a $\sim$ 120 $ns$ bunches support the required time structure of the beam pulses required for the Muon g-2 experiment. 

\begin{figure}[h]
\centering
\includegraphics[width=76mm]{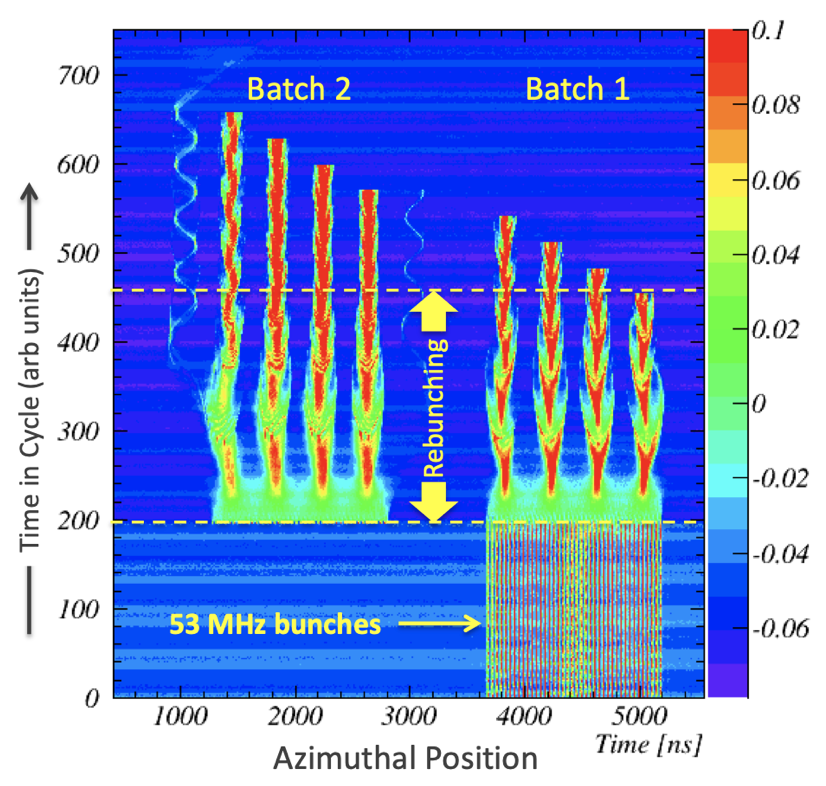}
\caption{Adiabatic rebunching in Recycler synchrotron.} \label{fig4}
\end{figure}

\subsection{Beam Delivery Performance Enhancement}
\begin{figure}[h]
\centering
\includegraphics[width=76mm]{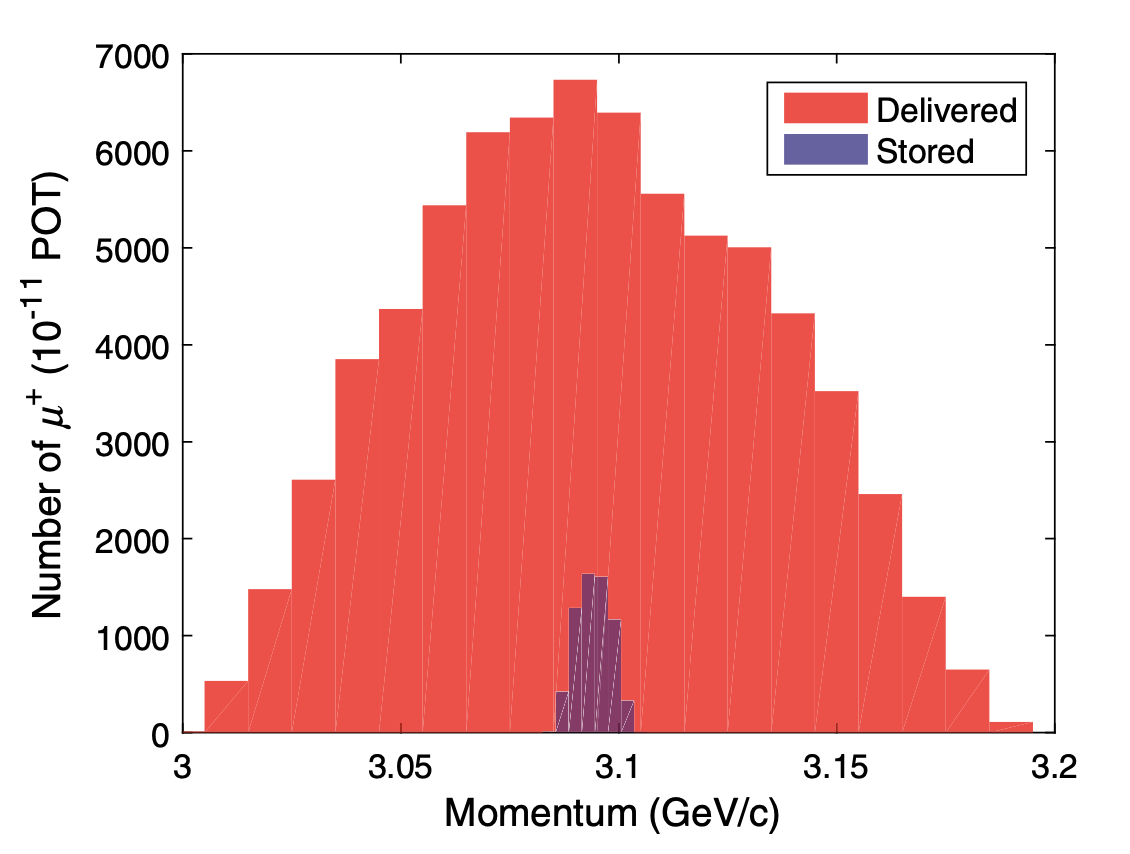}
\caption{Distribution of momentum of the Muon Campus beam at the Muon g-2 storage ring entrance from numerical simulations shown in red and momentum distribution of the same beam after 100 turns in the storage ring shown in blue. A large fraction of the incoming beam is lost and the RMS momentum spread of the remaining beam after 100 turns is $0.12\%$. POT refers to the  protons on target. Courtesy: Diktys Stratakis.} \label{fig5}
\end{figure}
\begin{figure}[h]
\centering
\includegraphics[width=76mm]{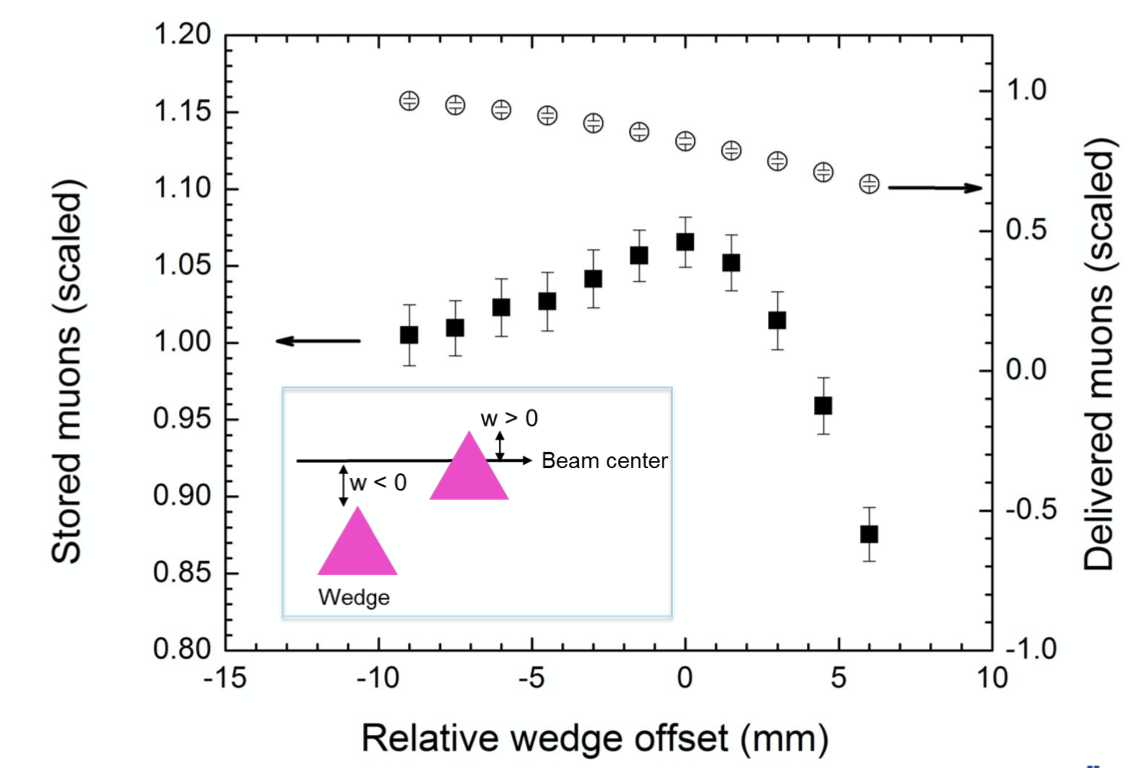}
\caption{Result from a proof-of-principle test: a boron carbide wedge provided a $7\%$ gain in stored muons. Courtesy: Diktys Stratakis.} \label{fig6}
\end{figure}

A polarized beam of muon is created by capturing forward muons from pion decays in flight and is injected into the Muon g-2 storage ring. From numerical simulation, it was shown that the acceptance of the muon delivery beamline is $1.26\%$ whereas only muons with rms momentum spread of $0.12\%$ would survive after about 100 turns in the storage ring of the experiment as shown in Figure \ref{fig5}. In order to store $\pm 5\%$ magic momentum muons in the storage ring, a scheme was used to reduce momentum spread to enhance the storage efficiency via separating the particles based on their momentum by guiding them into a dispersive area and then passing the beam through a wedge absorber \cite{wedge}. The system of the wedge absorber was installed and commissioned supported by Fermilab's LDRD. With a properly designed wedge, the high energy muons will lose more energy than the low energy muons as a result of which the overall energy spread of the beam will be reduced. It was shown that the number of stored muons was increased and the statistical uncertainty of the measurement could be minimized as a consequence by placing two wedges in the M5 beamline. A proof-of-principle test showed up to a $7\%$ improvement on the number of stored muons with a boron carbide wedge as shown in Figure \ref{fig6}.

\subsection{Muon g-2 Status}
The Muon g-2 experiment has recently published the very exciting run 1 result that  measured the anomalous magnetic moment ($\omega_a$) of positive muons to an unprecedented 460 ppb precision. Commissioning of the experiment took place between June 2017 and March 2018 which involved tuning injection parameters e.g., inflector, quad scraping, kicker settings \& radial field. Commissioning led to the first observation of processions of muons. Run 1 physics data was collected between March 2018 and June 2018. The Muon g-2 experiment has recently finished run 5 physics data, with total integrated statistics of $19\times$ BNL data. 

\subsection{Mu2e Beam Delivery Details}
\begin{figure}[h]
\centering
\includegraphics[width=76mm]{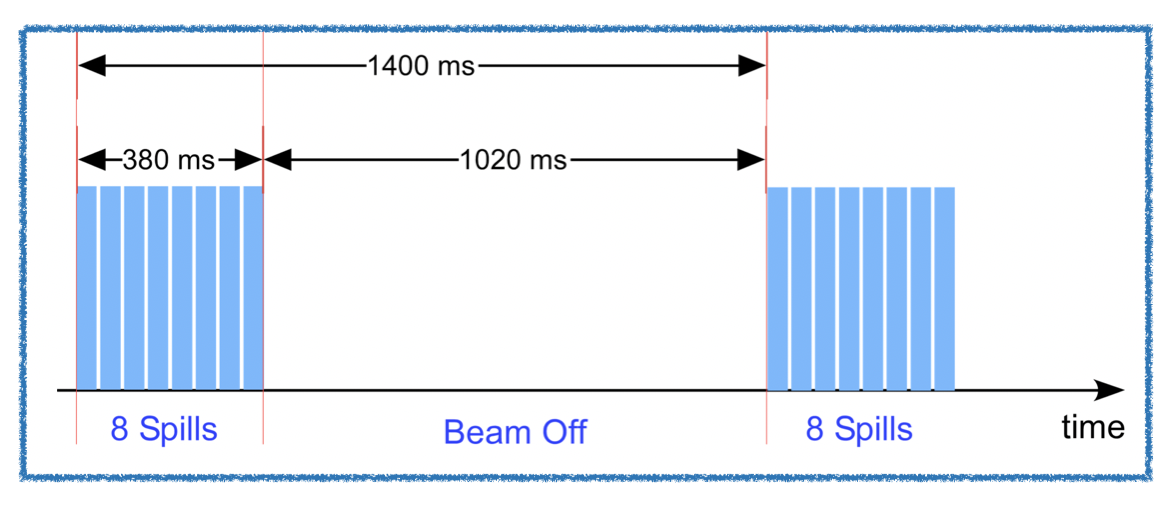}
\caption{Time structure of beam pulses to Mu2e experiment} \label{fig7}
\end{figure}

\begin{figure}[h]
\centering
\includegraphics[width=76mm]{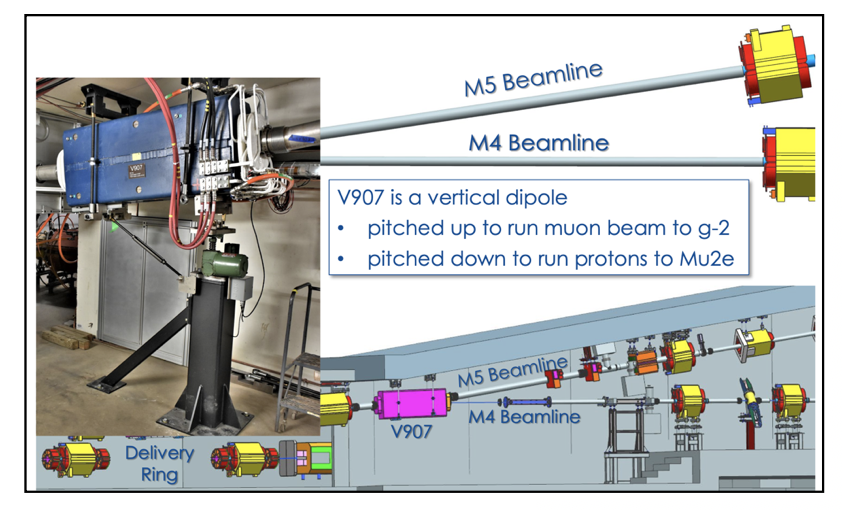}
\caption{Time structure of beam pulses to Mu2e experiment} \label{fig8}
\end{figure}

The signal that Mu2e is designed to detect is the coherent conversion of a muon into an electron in the field of an Aluminum (Al) nucleus. The beam preparation for the mu2e conversion is based on the slow resonant extraction \cite{mu2eresonant} of protons from the delivery ring. Mu2e uses a pulsed proton beam so that we get pulses of a secondary muon beam that is captured in Al atom and muon knocks off an electron out off the orbit, replaces the electron and orbits around in a bound state in an Al atom. While waiting for this muon to decay, another muon should not come in impinging on the Al atom. Hence we require the pulsed beam structure. Each of the 8 GeV proton bunch is synchronously transferred from the Recycler to the DR where the beam is resonantly extracted to Mu2e proton target via M4 line over 380 ms as shown in Figure \ref{fig7}. Each beam spill is 43 ms long with a 5 ms reset between two consecutive spills. Resulting 8 bunches are extracted every 48 ms. After the 8th spill, the Recycler is used for NuMI$/$NO$\nu$A slip-stacking. 
A vertical dipole shown in Figure \ref{fig8} is rotated and pitched down to send beam to Mu2e via M4 line instead of to the Muon g-2 experiment via M5 line when the vertical dipole is pitched up.
\subsection{Resonant extraction of beam for Mu2e}
\begin{figure}%
    \centering
    {{\includegraphics[width=76mm]{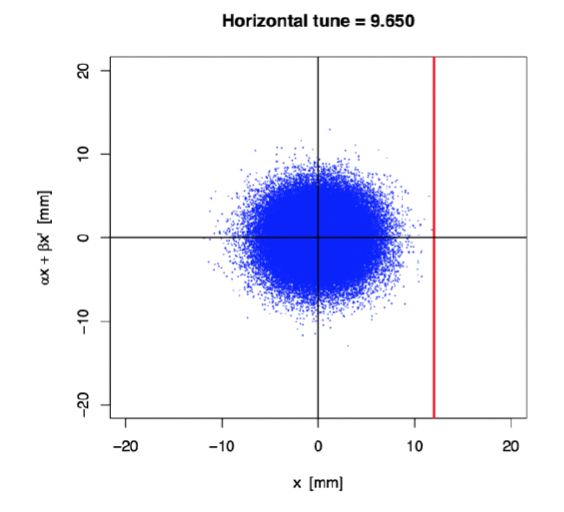} }}%
    \qquad
    {{\includegraphics[width=76mm]{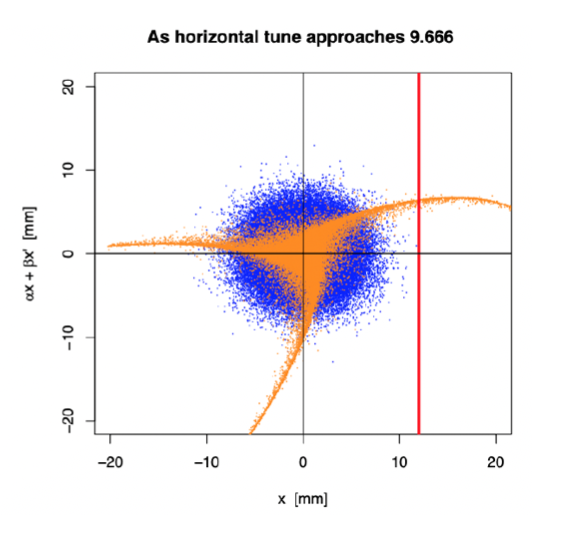} }}%
    \caption{Evolution of particles in a normalized phase-space shown as a blue circular blob (top), evolution of particles as the tune of the machine approaches the third integer resonance shown as the orange triangle (bottom).}%
    \label{fig:9}%
\end{figure}

The size of a particle beam is denoted by a property known as emittance which is roughly 
the area or volume in the phase space of the particles. The phase space variables in each spatial direction for a particle are the position and momentum components of the particle, denoted by x, $p_x$, y, $p_y$, z, and $p_z$ where time is an independent variable. So, the phase space plot characterizes a specific location in a machine as the positions and the momenta of the particle advance as the beam propagates. The horizontal phase space is defined by x (position) vs $p_x$ (transverse momentum component in x direction). When the transverse momenta are small compared to the longitudinal momentum components; $p_z >> p_x, p_y$, the transverse angles with respect to the ideal trajectory of the particle denoted by $x$, $y$ are used instead of the transverse momenta. If a particle accelerator consists only of dipole and quadrupole elements, it is considered a 'linear' system. For a linear system, the phase-space evolution of a single particle plotted in the ($x, x$) coordinates will be contained within an elliptical shape, even though the particle jumps from one point to another as it traverses through many quadrupole and dipole elements. When the evolution of the particles are plotted in the 'normalized' phase-space in the (x vs $\alpha x + \beta x$) coordinates, the particle will traverse a circle instead of an ellipse as shown in the Figure \ref{fig:9}, top plot. 
However, non-linearity in the system can be introduced by adding sextupole in the beam lattice. As the sextupole strength is ramped up, the circular blob in the normalized phase-space starts turning into a triangular distribution as shown in the Figure \ref{fig:9} bottom plot. This is because the sextupole suddenly introduces a triangular separatrix which defines the area within which the trajectories are stable. And the particles inside the separatrix keep evolving contained inside the phase-space and the particles outside the separatrix start to increase in amplitude by every turn.

The area of the stable region (i.e., the area of the triangle formed by the separatrices) depends not only on the sextupole strength but also how close the tune of the beam is to the third integer resonance. If the machine tune is far away from the third integer resonance, then triangle appears to be big, and almost all particles within the triangle are stably zooming around in circles in the normalized phase-space. But as the tune is driven closer to N=3 (where N is an integer), the triangle becomes smaller and the particles become unstable. Their x starts to increase, and once its value goes past the septum location value as indicated by the red vertical line in Figure \ref{fig:9} bottom, the particles get an electrostatic kick on to another beam line and are extracted. Choosing third integer resonance helps keep the losses minimum at the extraction point. This beam extraction technique of blowing up the beam in the horizontal direction by ramping up the quadrupole current and driving the beam tune close to third integer by introducing sextupole is known as the Resonant extraction. 
The main goal of the resonant extraction for the Mu2e experiment is to start with $\sim 10^{12}$ protons in the Delivery Ring and drive its tune from 9.650 to 9.666 in such a way that the area under the triangular separatrix is reduced in a delicate manner so that the blob in the phase-space shown in Figure \ref{fig:9} is blown up and $\sim 3X10^7$ protons are extracted that jump past $x_{sept} = 12 mm$ every turn. This is done over 25,000 turns (or 43 ms). 
The resonant extraction delivers the bunch structure required for the Mu2e experiment as shown in Figure \ref{fig:10}. 
The proton beam used is 250 ns wide micro bunch of protons that arrives every 1695 ns apart. 

\begin{figure}[h]
\centering
\includegraphics[width=76mm]{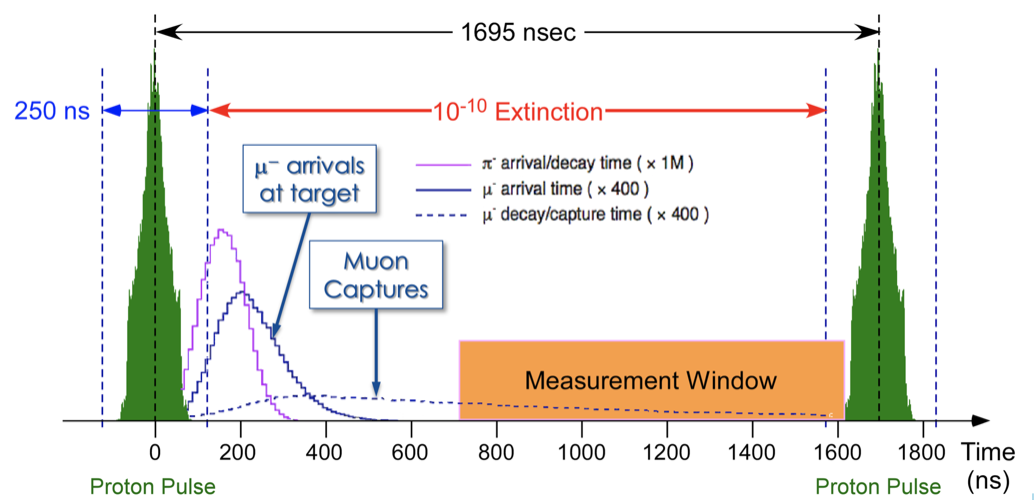}
\caption{Mu2e bunch structure.} \label{fig10}
\end{figure}
Beam related backgrounds can arise from pions, antiprotons and electrons that remain in the beam when the muon arrives at the stopping Al target. So, this beam structure helps reduce beam related background. 
Because out-of-time protons could produce prompt backgrounds, it is critical that there should be nothing between two consecutive proton bunches. This is referred to as “extinction” with  $< 1$ proton every 250 pulses. Extinction system makes sure that no proton comes to the production target between micro bunches.

\subsection{Beam Commissioning to the Mu2e Diagnostic Absorber}
Beam to the Diagnostic Absorber was run on April 14th, 2022 for the first time in the Mu2e beamline. Along the M4 beamline, multi-wires were placed at different strategic locations to monitor the beam profiles. 
\begin{figure}[h]
\centering
\includegraphics[width=76mm]{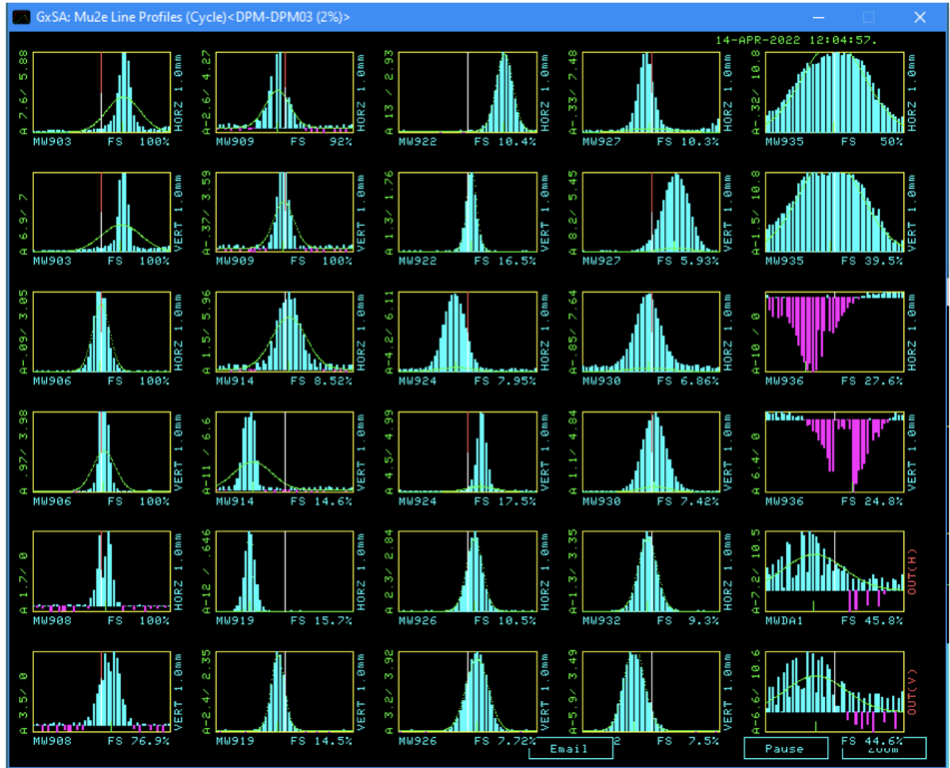}
\caption{Beam profiles in multi-wires placed along M4 line, upto Diagnostc Absorber.} \label{fig11}
\end{figure}
Figure \ref{fig11} shows the beam profile data collected from all these monitors. The fact that we see beam to all of them suggests that we had sent beam to the Diagnostic Absorber successfully. A toroid in the Diagnostic Absorber line will be placed in future to monitor beam intensity.
\subsection{Machine Learning Spill regulation}
The beam delivery from the Delivery Ring to the Mu2e experiment is achieved by the slow resonant extraction technique by driving the machine tune close to the third integer resonance. The strength of the resonance is controlled by two sextupole magnet circuits Driving the tune to an exact resonance value is controlled by three tune ramping quadrupoles via a Spill Regulation System (SRS). The SRS will regulate the slow extraction of the beam through the quadrupole circuit and the RF Knock-Out system. Slow changes in different accelerator components introduce instabilities in the proton beam in the DR. The SRS controller tracks the slow changes within a spill and provide correction to the current ramp needed to achieve the uniform spill rate. The REAL-TIME EDGE AI FOR DISTRIBUTED SYSTEMS (READS) \cite{ML} group is working on an end-to-end machine learning (ML) differentiable simulator parameterized by the PID gains to optimize the gains with simulated data. The Machine Learning simulator based on a neural network is used to find an optimal PID gain. With each training iteration, PID gains generated by the neural network are ingested by the PID simulator to produce a corrected spill.

\subsection{Unique proton target and remote handling}
Beam line components for the Mu2e experiment are made with as little material as possible to minimize particle interactions so as to reduce possible beam related backgrounds and reach the signal detection sensitivity required by the experiment. This led to designing the unique Mu2e production target \cite{mu2etarget} that looks like a thin pencil as shown in Figure \ref{fig:12}. The target needs to be suspended in a vacuum chamber inside a superconducting solenoid magnet. Hence the pencil-shaped target will be supported by a bicycle ring structure from which it will be suspended in vacuum. The target is radiatively cooled, i.e., it releases heat on its own, hence helps with reduction of beam power, cost, and safety concerns. Work is underway for application of robotics in remote handling for minimizing radiation exposure. The remote handling for the Mu2e production target will be a robotic machine that will travel from a side room to the target hall on floor rails. This robot will be able to remove and replace the target window and assembly from the production solenoid. 
\begin{figure}[h]
\centering
\includegraphics[width=76mm]{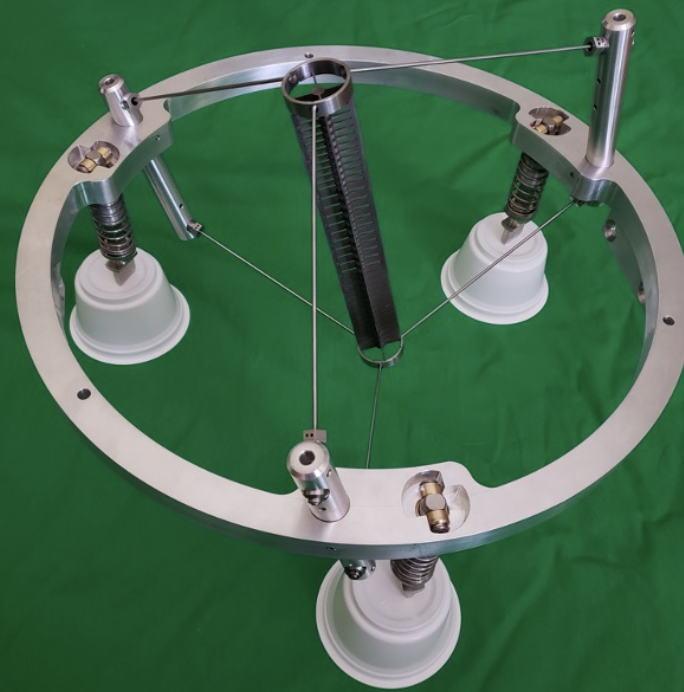}
\caption{Beam profiles in multi-wires placed along M4 line, up to Diagnostic Absorber.} \label{fig12}
\end{figure}

\subsection{Beyond Mu2e: Future of Muon Campus}
Beyond the Mu2e experiment, search for muon conversion in the PIP II \cite{pip2} era utilizing 100 kW beam to deliver 10 stopped muons in three years of data taking is a possibility by taking advantage of continuing detector improvements and accelerator upgrades at Fermilab. Besides Mu2e II \cite{mu2e2}, other potential experimental ideas \cite{workshop} \cite{agora} are being explored to build short-, medium-, and long-term muon- and non-muon-based experiments at Fermilab. 

\section{Summary}
An accelerator facility to provide beams to both the Muon g-2 and the Mu2e experiments has been designed and constructed at Fermilab. The facility has been commissioned in 2017 and is now in operation phase for Muon g-2 Experiment since 2018. It currently delivers roughly $1\times$ the BNL statistics per month. 
A first proof of principle of application of a passive wedge system in the beamline shows up to $7\%$ improvement on stored muons for the Muon g-2 experiments. Current plan is to run Muon g-2 until FY2023 while in parallel performing various commissioning tests for the Mu2e experiment beam delivery. 

\bigskip 

\end{document}